\documentclass[reprint, superscriptaddress, amsmath, amssymb, aps, prl]{revtex4-2}
\usepackage[dvipdfmx]{graphicx}
\usepackage{dcolumn}
\usepackage{bm}
\usepackage[dvipdfmx]{hyperref}
\usepackage{comment}
\usepackage{color}
\usepackage{physics}
\begin{document}
\preprint{APS/123-QED}
\title{Multi-locational Majorana Zero Modes}
\author{Yutaro Nagae}
\thanks{These authors contributed equally to this work.}
\affiliation{Department of Applied Physics, Nagoya University, Nagoya 464-8603, Japan}
\author{Andreas P. Schnyder}
\affiliation{Max-Planck-Institut f\"ur Festk\"orperforschung, Heisenbergstrasse 1, D-70569 Stuttgart, Germany}
\author{Yukio Tanaka}
\affiliation{Department of Applied Physics, Nagoya University, Nagoya 464-8603, Japan}
\author{Yasuhiro Asano}
\affiliation{Department of Applied Physics, Hokkaido University, Sapporo 060-8628, Japan}
\author{Satoshi Ikegaya}
\thanks{These authors contributed equally to this work.}
\affiliation{Department of Applied Physics, Nagoya University, Nagoya 464-8603, Japan}
\affiliation{Institute for Advanced Research, Nagoya University, Nagoya 464-8601, Japan}
\date{\today}

\begin{abstract}
We show the appearance of an unconventional Majorana zero mode whose wave function splits into multiple parts located at different ends of different one-dimensional topological superconductors,
hereinafter referred to as a multi-locational Majorana zero mode.
Specifically, we discuss the multi-locational Majorana zero modes in a three-terminal Josephson junction consisting of topological superconductors,
which forms an elemental qubit of fault-tolerant topological quantum computers.
We also demonstrate anomalously long-ranged nonlocal resonant transport phenomena caused by the multi-locational Majorana zero mode.
\end{abstract}
\maketitle

\textit{Introduction}.---%
Majorana zero modes (MZMs) of topological superconductors (TSs) have been a central issue in condensed matter physics%
~\cite{wilczek_09,kane_10,zhang_11,schnyder_16,flensberg_12,nagaosa_12,beenakker_13,sato_16,sato_17}.
This research field has made remarkable progress with finding various forms of the MZMs,
such as Majorana end modes in one-dimensional TSs~\cite{kitaev_01},
chiral MZMs in TSs with broken time-reversal symmetry~\cite{volovik_97,green_00,furusaki_01},
helical MZMs in time-reversal invariant TSs~\cite{maiti_06,schnyder_08,zhang_09,tanaka_09},
dispersionless MZMs in nodal TSs~\cite{buchholtz_81,nagai_86,tanuma_01,sato_11,schnyder_11},
and Majorana hinge/corner modes in higher-order TSs~\cite{balents_15,hughes_17,fang_17,brouwer_17,khalaf_18,volovik_10,zhu_18,wang_18,ikegaya_21}.

\begin{figure}[bbbb]
\begin{center}
\includegraphics[width=0.5\textwidth]{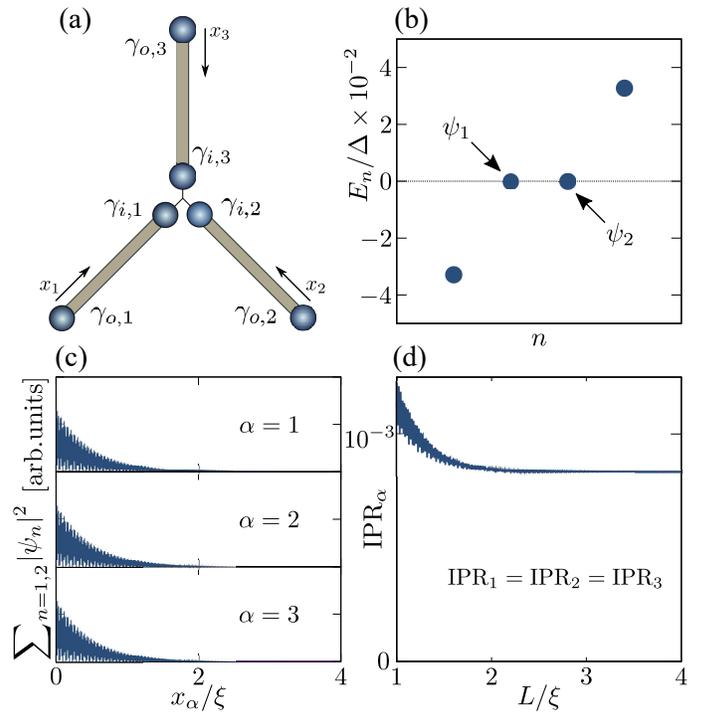}
\caption{(a) Schematic image of the topological TJJ.
(b) Energy eigenvalues of the topological TJJ, where all TSs have the same length, $L=800 \approx 4\xi$.
(c) Amplitudes of the MMM in each TS.
(d) Inverse participation ratio of the MMM in $\alpha=1$.}
\label{fig:figure1}
\end{center}
\end{figure}

On the basis of the bulk-boundary correspondence, which is an essential concept in the physics of topological condensed matters,
a MZM is usually localized at  \emph{one} end/surface of \emph{one} TS~\cite{kane_10,zhang_11,schnyder_16,thouless_82,kohmoto_85,hatsugai_02}.
In this Letter, nevertheless, we predict the appearance of an unconventional MZM whose wave function splits into \emph{multiple} parts located at \emph{different} ends of \emph{different} TSs,
where the distance between the ends hosting the MZM can exceed the decay length of the MZM substantially (see Fig.~\ref{fig:figure1}).
Hereinafter, we refer to this particular zero mode, which exhibits simultaneously the Majorana nature and the nonlocal nature, as a multi-locational Majorana zero mode (MMM).
Specifically, we discuss the appearance of the MMM in a three-terminal Josephson junction (TJJ) consisting of one-dimensional TSs (hereinafter, referred to as a topological TJJ).

Another important topic in the field of MZMs is the transport anomalies in junctions consisting of topological superconductors.
So far, the following transport signatures originating from MZMs have been investigated: 
zero-bias conductance quantization in normal-metal--superconductor junctions~\cite{tanaka_95,sarma_01,tanaka_04,tanaka_05(1),law_09,asano_13,ikegaya_15,ikegaya_16(1)},
fractional current-phase relationship in Josephson junctions~\cite{tanaka_97,kwon_04,asano_06(1),ikegaya_16(2)},
anomalous transport in multi-terminal devices~\cite{beenakker_08,pachos_10,law_13,ardonne_17,sen_18,martin_19,soori_19,asano_20,flensberg_20,soori_20,houzet_21,%
law_14,akhmerov_18,melin_11,shtrikman_18,girit_19,schmidt_14,lutchyn_15,beenakker_15(2)},
and quantum teleportation in devices including capacitors~\cite{fu_10,egger_11,egger_12}.
In this Letter, we demonstrate that the MMM causes striking nonlocal transport phenomena, which manifests clearly the coexistence of the Majorana nature and the nonlocal nature.

The topological TJJ forms an elemental qubit of fault-tolerant topological quantum computers%
~\cite{ivanov_01,kitaev_03,sarma_08,alicea_16,fisher_11,tewari_11,sarma_15,sen_16,sakurai_20,dagotto_23}.
At present, topological TJJs have not yet been realized experimentally.
However, on the way to realizing the topological quantum computations,
which is a central goal of the topological condensed matter physics, future experiments will certainly build the topological TJJs.
The observations of the characteristic transport phenomena due to the MMM provide smoking-gun evidences for the realizations of fully functional topological TJJs.

\textit{Multi-locational Majorana zero modes}.---%
At first, we consider a low-energy effective Hamiltonian describing the couplings between MZMs in a topological TJJ.
As illustrated in Fig.~\ref{fig:figure1}(a), in the absence of any couplings, one MZM appears at each end of the three TSs,
where we denote the Majorana operator for the outer (inner) MZM of the $\alpha$-th TS with $\gamma_{o(i),\alpha}=\gamma^{\dagger}_{o(i),\alpha}$.
In the TJJ, the couplings among these MZMs are described by
\begin{align}
\begin{split}
&H = \frac{1}{2} \boldsymbol{\gamma}^{\mathrm{T}} \mathcal{H} \boldsymbol{\gamma},\quad
\mathcal{H}=\left( \begin{array}{cc} 0_{3\times3} & i \mathcal{H}_T \\ - i \mathcal{H}_T & \mathcal{H}_J \end{array} \right),\\
&\mathcal{H}_T = \mathrm{diag}[E_{T,1},E_{T,2},E_{T,3}],\\
&\mathcal{H}_{J} = \left( \begin{array}{ccc} 0 & i E_{J,12} & i E_{J,13} \\  -i E_{J,12} & 0 & i E_{J,23}\\ -i E_{J,13} & -i E_{J,23} & 0 \end{array} \right),
\label{eq:ham}
\end{split}
\end{align}
where $\boldsymbol{\gamma}=[\gamma_{o,1},\gamma_{o,2},\gamma_{o,3},\gamma_{i,1},\gamma_{i,2},\gamma_{i,3}]^{\mathrm{T}}$.
The $3\times3$ null matrix is represented by $0_{3\times3}$. 
The couplings between the outer and inner MZMs of the $\alpha$-th TS are characterized by $E_{T,\alpha} \propto \exp (-2L_\alpha/\xi)$~\cite{sarma_08,stanescu_12},
where $L_\alpha$ and $\xi$ represent the length of the $\alpha$-th TS and the decay length of the MZM, respectively.
The couplings among the inner MZMs at the Josephson junction is described by $\mathcal{H}_J$~\cite{kitaev_01},
where $E_{J,\alpha\beta} =C_{\alpha\beta} \sin [(\phi_{\alpha}-\phi_{\beta})/2]$, with $C_{\alpha\beta}=C_{\beta\alpha}$ and $\phi_{\alpha}$ representing the superconducting phase of the $\alpha$-th TS.
For simplicity, we ignore long-range couplings, i.e., $\gamma_{o,\alpha}$ couples only with $\gamma_{i,\alpha}$.

Since originally there were three MZMs at the Josephson junction, $H_J$ becomes a $3\times3$ skew-symmetric matrix, which always has an unpaired zero-energy eigenvalue.
The two remaining finite eigenvalues are given by $\pm E_J$ with $E_J = \sqrt{E_{J,12}^2 + E_{J,23}^2 + E_{J,31}^2}$~\cite{sen_16,sakurai_20}.
By applying a unitary transformation to diagonalize $\mathcal{H}_{J}$, as also discussed in the Supplemental Material (SM)~\cite{sm}, we rewrite the Hamiltonian as
\begin{align}
\begin{split}
&H = \frac{1}{2} \tilde{\boldsymbol{\gamma}}^{\dagger} 
\left( \begin{array}{cc} \mathcal{H}_{\mathrm{eff}} & \mathcal{V} \\ \mathcal{V}^{\dagger} & \mathcal{E}_J \end{array} \right) \tilde{\boldsymbol{\gamma}},\\
&\mathcal{H}_{\mathrm{eff}} = \left( \begin{array}{cccc} 0 & 0 & 0& iv_{1} \\ 0 & 0 & 0 & i v_{2} \\  0 & 0 & 0 & i v_{3}\\ -iv_{1} & -i v_{2} & -i v_{3} & 0 \end{array} \right),\\
&\mathcal{V}=\left(\begin{array}{cc} i v_{1,+} & iv_{1,-} \\ iv_{2,+} & iv_{2,-} \\ iv_{3,+} & iv_{3,-} \\ 0 & 0\end{array} \right),\quad
\mathcal{E}_J=\left(\begin{array}{cc}E_J & 0 \\ 0 & -E_J\end{array} \right),
\end{split}
\end{align}
where $\tilde{\boldsymbol{\gamma}} = [\gamma_{o,1},\gamma_{o,2},\gamma_{o,3},\gamma_{0,J},\gamma_{+,J},\gamma_{-,J}]^{T}$,
with $\gamma_{0,J}$ and $\gamma_{\pm,J}$ corresponding to the zero-energy and the finite-energy eigenstates of $\mathcal{H}_J$, respectively.
Specifically, we obtain $v_{\alpha}= (E_{T,\alpha}/2) \sum_{\beta,\gamma}\epsilon^{\alpha\beta\gamma}( E_{J,\beta\gamma}/E_J)$, where $\epsilon^{\alpha\beta\gamma}$ is the Levi-Civita symbol.
The components in $\mathcal{V}$ are also of the order of $E_{T,\alpha}$~\cite{sm}.
Thus, in the limit of $E_{T} \ll E_J$ with $E_T=\sqrt{E_{T,1}^2+E_{T,2}^2+E_{T,3}^2}$, the low-energy excitation of $|E| \ll E_J$ is approximately described by~\cite{fu_18,timm_18},
\begin{align}
\begin{split}
&H_{\mathrm{eff}} = \frac{1}{2} \boldsymbol{\gamma}_{\mathrm{eff}}^{T} \mathcal{H}_{\mathrm{eff}} \boldsymbol{\gamma}_{\mathrm{eff}},\\
&\boldsymbol{\gamma}_{\mathrm{eff}} = [\gamma_{o,1},\gamma_{o,2},\gamma_{o,3},\gamma_{0,J}]^{T},
\label{eq:heff}
\end{split}
\end{align}
where the contributions from the finite-energy eigenstates at the Josephson junction (i.e., $\gamma_{\pm,J}$) are ignored.
The detailed derivation of $H_{\mathrm{eff}}$ is presented in the SM~\cite{sm}.
The effective low-energy Hamiltonian $H_{\mathrm{eff}}$ has doubly degenerate zero modes:
\begin{align}
\gamma_{n} = \sum_{\alpha=1\text{-}3} c_{\alpha}^{n} \gamma_{o,\alpha} \quad (n=1,2),
\label{eq:mmm_general}
\end{align}
where the coefficients $c_{\alpha}^{n}$ are real numbers satisfying
\begin{align}
\sum_{\alpha}v_{\alpha}c_{\alpha}^{n}=0, \quad \sum_{\alpha}c_{\alpha}^{n}c_{\alpha}^{m}=\delta_{m,n}.
\end{align}
Importantly, since the coefficients $c_{\alpha}^{n}$ are \emph{real} numbers, the Majorana relation holds, i.e., $\gamma_{n}=\gamma_{n}^{\dagger}$.
Moreover, $\gamma_n$ is described by the superposition of the outer end MZMs (i.e., $\gamma_{o,\alpha}$).
Namely, the wave function of $\gamma_n$ has large amplitudes at the outer edges of different TSs.
This unconventional zero mode exhibiting both Majorana nature and the nonlocal nature is what we refer to as the MMM.
Due to the degeneracy at zero energy, there is ambiguity in the coefficients $c_{\alpha}^{n}$.
However, the quantity of
\begin{align}
\begin{split}
&P_{\alpha} = \sum_{n=1,2}|c_{\alpha}^{n}|^2=\sum_{\beta\neq\alpha}\frac{v^2_{\beta}}{E_v^2},\\
&E_v=\sqrt{v_1^2+v_2^2+v_3^3},
\end{split}
\end{align}
which characterizes the amplitudes of the MZMs in the $\alpha$-th TS, is independent of the choice of $c_{\alpha}^{n}$.
As long as $v_{\alpha=1,2,3}\neq0$, we obtain $P_{\alpha}<1$ for all TSs.
Namely, \emph{two} MMMs are distributed over the ends of \emph{three} different TSs.
For instance, we obtain $P_1=P_2=P_3=2/3$ with $v_1=v_2=v_3$, which means that the \emph{two} MMMs are equality distributed over the ends of \emph{three} different TSs.
When one of $v_{\alpha}$ is zero, e.g. $v_1=v_2\neq0$ and $v_3=0$, we obtain $P_1=P_2=1/2$ and $P_3=1$.
In this case, \emph{one} MMM is distributed over the outer edges of \emph{two} TSs, namely the first and second one, while the third TS hosts a conventional MZM at its outer edge.

Here we numerically reproduce the appearance of the MMM by using the tight-binding Bogoliubov-de Gennes (BdG) Hamiltonian of a spin-less $p$-wave superconductor (i.e., a Kitaev chain~\cite{kitaev_01}).
We note that the low-energy physics of various one-dimensional TSs~\cite{stern_21,yacoby_23},
such as superconducting semiconductor nanowires~\cite{sarma_10,oreg_10,kouwenhoven_12,deng_12},
magnetic atom chains deposited on superconductors~\cite{beenakker_11,yazdani_13,yazdani_14,yazdani_17},
and planer topological Josephson junctions~\cite{flensberg_17,halperin_17,haim_19,setiawan_19,nichele_19,yacoby_19,shabani_20,shabani_21,ikegaya_22},
are described effectively by the Kitaev chain~\cite{asano_13,beenakker_11,ikegaya_22}.
The BdG Hamiltonian reads $H_{\mathrm{BdG}}= \sum_{\alpha=1\text{-}3}H_{\alpha} + H_d$ with
\begin{align}
H_{\alpha} =& \sum_{x_{\alpha}=1}^{L_{\alpha}-1} ( -t c^{\dagger}_{x_{\alpha}+1}c_{x_{\alpha}} + \mathrm{h.c.})
- \sum_{x_{\alpha}=-\infty}^{L_{\alpha}} \mu c^{\dagger}_{x_{\alpha}}c_{x_{\alpha}} \nonumber\\ 
&+\frac{1}{2}\sum_{x_{\alpha}=1}^{L_{\alpha}-1}(i\Delta e^{i\phi_{\alpha}} c^{\dagger}_{x_{\alpha}+1}c^{\dagger}_{x_{\alpha}} + \mathrm{h.c.}),\\
H_d =& \sum_{\alpha=1\text{-}3} (-t_{d,\alpha} c^{\dagger}_{d}c_{L_{\alpha}} + \mathrm{h.c.}) - \mu_{d} c^{\dagger}_{d}c_{d}, \nonumber
\end{align}
where $c_{x_{\alpha}}$ ($c^{\dagger}_{x_{\alpha}}$) is the annihilation (creation) operator of an electron at a site $x_{\alpha}$ of the $\alpha$-th branch,
$t$ is the nearest-neighbor hopping integral, $\mu$ denotes the chemical potential, and $\Delta$ represents the pair potential in the superconducting segment.
The three TSs are connected through a single normal site, where the coupling is described by $H_d$.
In the following calculations, we fix the parameters as $\mu=\mu_d=-0.8t$, $t_{d,\alpha}=t$, and $\Delta=0.01t$,
where the decay length of MZMs is evaluated by $\xi = 1/\mathrm{artanh} (\Delta/2t) \approx 200$.
For simplicity, we assume that all TSs have a same length $L_1=L_2=L_3=L$.
In Fig.~\ref{fig:figure1}(b), we show the energy eigenvalues obtained by diagonalizing $H_{\mathrm{BdG}}$, where $L=800 \approx 4\xi$.
The superconducting phases are chosen as $(\phi_1,\phi_2,\phi_3)=(0,2\pi/3,4\pi/3)$, which corresponds to $v_1=v_2=v_3$ for the effective Hamiltonian $H_{\mathrm{eff}}$.
We observe two zero-energy states, marked by the arrows, corresponding to two MMMs.
The first excited states are approximately located at $E=\pm E_v$, which characterizes the necessary energy resolution to observe the MMMs.
The energy $E_v$ is proportional to $E_{T,\alpha}\approx \Delta \cos(k_FL) e^{-2L_{\alpha}/\xi}$ with $k_F=\arccos(\mu/2t)$ and thus oscillates and decays by increasing $L$ ~\cite{stanescu_12}.
We note that these conditions might be achievable in experiments with artificial TSs%
~\cite{sarma_10,oreg_10,kouwenhoven_12,deng_12,beenakker_11,yazdani_13,yazdani_14,yazdani_17,
flensberg_17,halperin_17,haim_19,setiawan_19,nichele_19,yacoby_19,shabani_20,shabani_21,ikegaya_22},
where the coupling strength $E_{T, \alpha}$ as well as the decay length of the MZM can be controlled, for example, by changing the applied Zeeman fields~\cite{stanescu_12}.
In Fig.~\ref{fig:figure1}(c), we show the amplitude of the wave function of the zero energy states in each TS, i.e.,
$\sum_{n=1,2}|\psi_n(x_\alpha)|^2$ with $\psi_n(x_\alpha)$ representing the wave function of the $n$-th zero-energy states.
The amplitude of the wave function increases at the outer edge of each TS and is equivalent for all TSs,
which agrees with $P_1=P_2=P_3=2/3$ obtained by the effective theory based on $H_{\mathrm{eff}}$.
In Fig.~\ref{fig:figure1}(d), we show the inverse participation ratio~\cite{thouless_74} for each TS,
defined by $\mathrm{IPR}_{\alpha} = \sum_{n=1,2} \sum_{x_{\alpha}=1}^{L_{\alpha}}|\psi_n(x_\alpha)|^4$.
By increasing the length of TS (i.e., $L$), $\mathrm{IPR}_{\alpha}$ becomes a constant,
which means that the wave function of the zero-energy states is well localized at the outer edges of the TS and is not distribute in the entire system.
As a result, we numerically verified the appearance of the MMM in the tight-binding model.

\textit{Transport anomalies}.---%
We study transport anomalies due to the MMM.
At first, we capture the essence of transport properties of MMM by an analytical calculation using the coupling Hamiltonian in Eq.~(\ref{eq:ham}).
Then, we numerically reproduce the analytical results by employing recursive Green's function techniques~\cite{fisher_81,ando_91}, where the TSs are described by the Kitaev chains~\cite{kitaev_01}.
To study the transport properties of the MMM, we attach a normal-metal lead to the outer end of each TS.
We refer to the lead attached to the $\alpha$-th TS as the $\alpha$-th lead.
In general, for the wide-bandwidth normal-metal leads, the scattering matrix at energy $E$ are represented by~\cite{beenakker_08,flensberg_20}
\begin{align}
\begin{split}
S(E) &=\left( \begin{array}{cc}s^{ee} & s^{eh} \\ s^{he} & s^{hh} \end{array} \right)\\
&= 1 - 2 \pi i W^\dagger \left[ E - \mathcal{H} + i \pi WW^\dagger \right]^{-1} W,
\end{split}
\end{align}
where $(s^{\eta\zeta})_{\alpha\beta} = s^{\eta\zeta}_{\alpha\beta}$
with $s^{ee}_{\alpha\beta}$ and $s^{he}_{\alpha\beta}$ ($s^{eh}_{\alpha\beta}$ and $s^{hh}_{\alpha\beta}$)
being the scattering coefficients from the electron (hole) in the $\beta$-th lead to the electron and hole in the $\alpha$-th lead, respectively.
In our case, the coupling matrix $W$ is given by~\cite{beenakker_08}
\begin{align}
W=\left( \begin{array}{cc} w & -w \\ 0_{3\times3} & 0_{3\times3} \end{array} \right), \quad
w=\left( \begin{array}{ccc} w_1 & 0 & 0 \\ 0 & w_2 & 0 \\ 0 & 0 & w_3 \end{array} \right), 
\end{align}
where the basis for the normal-metal leads is chosen as
$\boldsymbol{c}=[c_{e,1},c_{e,2},c_{e,3},c_{h,1},c_{h,2},c_{h,3}]^{\mathrm{T}}$ with $c_{e(h),\alpha}$ representing a propagating electron (hole) in the $\alpha$-th lead.
$w_{\alpha}$ characterizes the couplings between the electron/hole in the $\alpha$-th lead and the MZM at the outer end of the $\alpha$-th TS.
Without loss of generality, we set $w_\alpha$ to be real numbers~\cite{beenakker_08}.
When $w_{\alpha}^2 \ll E_{T} \ll E_J$, the scattering coefficients at zero energy is obtained by
\begin{align}
\begin{split}
&s^{ee}_{\alpha\alpha} = \frac{v_{\alpha}^2 w_{\beta}^2 w_{\gamma}^2}{A},\quad
s^{he}_{\alpha\alpha} = -\frac{w_{\alpha}^2(v_{\beta}^2 w_{\gamma}^2+w_{\beta}^2 v_{\gamma}^2)}{A},\\
&s^{ee}_{\alpha\beta}=-s^{he}_{\alpha\beta}=\frac{v_{\alpha}v_{\beta}w_{\alpha}w_{\beta}w^{2}_{\gamma}}{A},\\
&A=v_1^2w_2^2w_3^2+v_2^2w_3^2w_1^2+v_3^2w_1^2w_2^2,
\label{eq:smat}
\end{split}
\end{align}
where $\alpha\neq\beta$, $\beta\neq\gamma$, and $\gamma\neq\alpha$.
The detailed derivation for the scattering coefficients are shown in the SM~\cite{sm}.
The scattering coefficients show highly intertwined properties:
the local Andreev reflection probability in the $\alpha$-th lead is related with the normal reflection probabilities in the different leads as
$|s^{he}_{\alpha\alpha}|=|s^{ee}_{\beta\beta}|+|s^{ee}_{\gamma\gamma}|$;
moreover, all nonlocal scattering probabilities between the $\alpha$-th and $\beta$-th leads are equivalent
(i.e., $|s^{ee}_{\alpha\beta}|=|s^{he}_{\alpha\beta}|=|s^{ee}_{\beta\alpha}|=|s^{he}_{\beta\alpha}|$).
When $v_1=v_2=v_3$ and $w_1=w_2=w_3$, for instance,
we obtain $|s^{he}_{\alpha\alpha}|^2=4/9$ and $|s^{ee}_{\alpha\alpha}|^2=|s^{ee}_{\alpha\beta}|^2=|s^{he}_{\beta\alpha}|^2=1/9$ for $\alpha \neq \beta$.
When $v_1=v_2$ and $v_3=0$ with $w_1=w_2=w_3$, we obtain $|s^{ee}_{\alpha\beta}|^2=|s^{he}_{\alpha\beta}|^2=1/4$ for $\alpha,\beta=1,2$.
Such highly intertwined scatterings with respect to the multiple leads are obviously due to the MMM.

\begin{figure}[tttt]
\begin{center}
\includegraphics[width=0.5\textwidth]{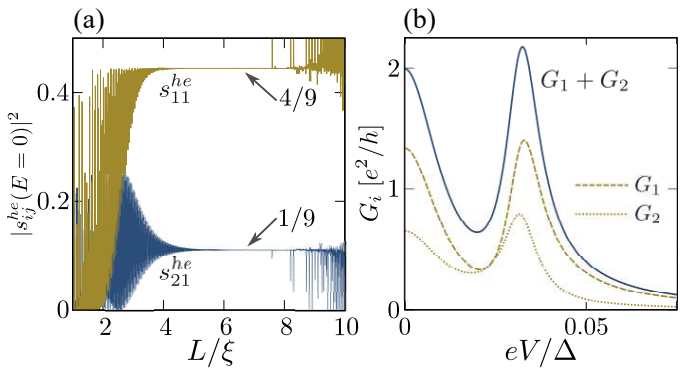}
\caption{(a) Nonlocal Andreev reflection probability $|s^{he}_{12}|^2$ and nocal Andreev reflection probability $|s^{he}_{11}|^2$ at zero energy as a function of the superconductor length.
(b) Differential conductance as a function of the bias voltage.}
\label{fig:figure2}
\end{center}
\end{figure}

We numerically reproduce the anomalous transport of the MMM by using the tight-binding BdG Hamiltonian.
We describe the normal leads located at $x_{\alpha}<1$ by setting $\Delta=0$ in $H_{\alpha}$
and denote the hopping integral at the interface of the $\alpha$-th lead and the $\alpha$-th TS (i.e., the hopping integral between $x_{\alpha}=0$ and $x_{\alpha}=1$) by $t^{\prime}_{\alpha}$.
We numerically compute the scattering coefficients using the recursive Green's function techniques~\cite{fisher_81,ando_91}.
In Fig.~\ref{fig:figure2}(a), we show the local Andreev reflection probability of $|s^{he}_{11}|^2$ and the nonlocal Andreev reflection probability of $|s^{he}_{12}|^2$
 at zero energy as a function of the superconductor length $L$.
We consider a low-transparency junction with $t_{\alpha}^{\prime}=0.1t$, which corresponds to $w_1=w_2=w_3$.
The superconducting phases are chosen as $(\phi_1,\phi_2,\phi_3)=(0,2\pi/3,4\pi/3)$, which corresponds to $v_1=v_2=v_3$.
We clearly find the plateau at $|s^{he}_{11}|^2=4/9$ and $|s^{he}_{12}|^2=1/9$ in a broad range of $L$, which agrees with Eq.~(\ref{eq:smat}).
We also confirm that other scattering amplitudes also exhibit the plateaus consistent with Eq.~(\ref{eq:smat}).
For longer $L$ exceeding $7\xi$, the ideal plateau is affected by the finite energy states at $E = \pm E_v \propto e^{-2L_{\alpha}/\xi}$. 
For $L$ shorter than $5\xi$, we see the deviation from the ideal plateau owing to more complex couplings neglected in $H_{\mathrm{eff}}$.
Nevertheless, it has been shown that typical nonlocal scatterings through superconductors decay exponentially
with the distance between the leads on a scale fixed by $\xi$~\cite{beenakker_08,law_13,law_14,akhmerov_18,flensberg_20,soori_20}.
On the other hand, in the topological TJJ, the plateau at $|s^{he}_{12}|^2=1/9$ is maintained even when the distance between the two leads (i.e., $2L$) exceeds $10\xi$
(the leads are separated by two TSs).
As a result, the robust plateau in Fig.~\ref{fig:figure2}(a) proves the anomalously long-ranged nonlocal transport mediated by the MMM.

We describe how the characteristic scatterings of the MMM affect the charge transport.
For this purpose, we apply the same bias voltage $V$ to all normal-metal leads, while all TSs are grounded.
We assume significantly low transparency at the interfaces between the leads and TSs, such that the bias voltages applied to the leads drop only at the interfaces.
We also assume $(\phi_1,\phi_2,\phi_3)=(0, 0, \pi)$, such that the Josephson currents between the TSs are absent;
this assumption yields $v_3=0$, and thus the MMM is distributed in the first and second TSs.
Within the Blonder--Tinkham--Klapwijik (BTK) formalism, the differential conductance $G_{\alpha}(eV) = d I_{\alpha}/d V$ at zero-temperature is given by~\cite{klapwijk_82,deutscher_00}
\begin{align}
G_{\alpha}(eV) = \frac{e^2}{h} \sum_{\beta} \left[\delta_{\alpha\beta} - |s^{ee}_{\alpha\beta}|^2 + |s^{he}_{\alpha\beta}|^2\right]_{E=eV},
\end{align}
where $I_{\alpha}$ is the time-averaged current in the $\alpha$-th lead.
We note that our calculations based on the BTK formalism is qualitatively justified for bias voltages well below the superconducting gap. 
The differential conductance at zero bias voltages with low transparency interfaces (i.e., $w_{\alpha}^2 \ll E_v \ll E_{J}$) is given by
\begin{align}
\begin{split}
&G_{1(2)}(0)=\frac{2e^2}{h}\frac{v_{2(1)}^2 w_{1(2)}^2}{v_1^2w_2^2+v_2^2w_1^2},\quad
G_{3}(0)=\frac{2e^2}{h}.
\end{split}
\end{align}
The differential conductance of the third lead exhibits the zero-bias conductance quantization~\cite{tanaka_95,sarma_01,tanaka_04,tanaka_05(1),law_09}.
On the other hand, $G_{1}(0)$ and $G_{2}(0)$ depend on the transparency of the interfaces and are not quantized.
Nevertheless, a remarkable relation holds:
\begin{align}
G_{1}(0)+G_{2}(0) = \frac{2e^2}{h}.
\label{eq:nqc}
\end{align}
Namely, the sum of zero-bias differential conductance in the different leads is perfectly quantized.
The observation of the nonlocally quantized conductance can be a smoking-gun evidence for the appearance of the MMM.
In Fig.~\ref{fig:figure2}(c), we show the differential conductance obtained numerically by using the tight-binding model.
We choose $L=800 \sim 4\xi$, $(\phi_1,\phi_2,\phi_3)=(0,0,\pi)$ and $(t_{1}^{\prime},t_{2}^{\prime},t_{3}^{\prime}) = (0.07t,0.1t,0.07t)$.
While $G_1\neq G_2$ owing to $t_{1}^{\prime} \neq t_{2}^{\prime}$, we see $G_1+G_2$ at zero-bias voltage is quantized to $2e^2/h$.
Therefore, we numerically demonstrate the sum quantization of the conductance due to the MMM.
We also find a second peak at the finite bias voltage $eV \approx E_v$.
We remark that the energy $E_v$, which characterizes the necessary resolutions in experiments, can be controlled, for instance, by changing applied Zeeman fields of the artificial TSs~\cite{stanescu_12}.

Even though the current shot noise is more difficult to measure experimentally than the differential conductance, it also manifests the profound nature of the intertwined scatterings of the MMM.
We specifically discuss the zero-frequency noise power defined by
\begin{align}
P_{\alpha\beta} = \int^{\infty}_{-\infty} \overline{\delta I_{\alpha}(0) \delta I_{\beta}(t)} dt,
\end{align}
where $\delta I_{\alpha}(t) = I_{\alpha}(t) - I_{\alpha}$ denotes the deviation of the current at time $t$ from the time averaged current.
Within the BTK formalism, the zero-frequency noise power at zero-temperature is given by~\cite{beenakker_94,datta_96}
\begin{align}
\begin{split}
&P_{\alpha\beta} = \frac{e^2}{h}\int^{eV}_{0} \mathcal{P}_{\alpha\beta}(E) dE,\\
&\mathcal{P}_{\alpha\beta}(E) = \delta_{\alpha\beta}\sum_{\eta=e,h}p^{\eta\eta}_{\alpha\alpha}
- \sum_{\eta,\zeta}\sigma_{\eta}\sigma_{\zeta}p^{\eta\zeta}_{\alpha\beta}p^{\zeta\eta}_{\beta\alpha},\\
&p^{\eta\zeta}_{\alpha\beta}=\sum_{\gamma=1,2,3}s^{\eta e}_{\alpha \gamma} (s^{\zeta e}_{\beta\gamma})^{\ast},
\end{split}
\end{align}
where $\sigma_{\eta}=1$ ($-1$) for $\eta=e$ ($h$).
For significantly low bias voltages with low-transparency interfaces, such as $eV \ll w_{\alpha}^2 \ll E_v \ll E_{J}$, the zero-frequency noise power is given by
\begin{align}
\begin{split}
&P_{11}=P_{22}=- P_{12} = - P_{21} = \frac{2h}{eV} I_1 I_2,\\
&P_{33}=P_{3\alpha}=P_{\alpha 3}=0, \quad (\alpha \neq 3),
\end{split}
\end{align}
where we approximate $P_{\alpha\beta} \approx (e^3V/h) \mathcal{P}_{\alpha\beta}(0)$ and $I_\alpha \approx G_{\alpha\alpha}(0)V$, which is justified in the linear response regime of $eV$.
We find that the particular relation,
\begin{align}
2|P_{12}|=P_{11}+P_{22},
\label{eq:noise1}
\end{align}
is satisfied.
The cross-correlator in any stochastic processes is bounded by the auto-correlator as $2|P_{12}| \leq P_{11}+P_{22}$.
Thus, the maximized cross-correlator of $|P_{12}|$ implies that the MMM causes the perfect correlation between the first and second leads.
In addition, the total noise power satisfies $\sum_{\alpha,\beta}P_{\alpha\beta}=0$,
which implies that the total charge current flowing into the TSs is noiseless due to the resonant transmissions of the conventional MZM at the third TS and the MMM splitting into the first and second TSs.
Especially, the relation of
\begin{align}
P_{11}+P_{22}+P_{12}+P_{21}=0
\label{eq:noise2}
\end{align}
is closely connected with the nonlocally quantized conductance in Eq.~(\ref{eq:nqc});
both are the physical consequence of the highly intertwined resonant scatterings caused by the MMM.
In the SM~\cite{sm}, we also demonstrate the anomalous current noise [i.e. Eq.~(\ref{eq:noise1}) and Eq.~(\ref{eq:noise2})] using the tight-binding model.

\textit{Discussion}.---%
In this paper, we discuss the charge currents only when the Josephson currents are absent, i.e., the superconducting phases are fixed at $(\phi_1,\phi_2,\phi_3)=(0,0,\pi)$.
Studying other cases, for example,
the coexistence of charge currents due to the bias voltages and Josephson currents~\cite{sen_18,martin_19,houzet_21}, remains as an important future task.
Nevertheless, we still expect drastic transport anomalies because as shown in Eq.~(\ref{eq:mmm_general}), the topological TJJ hosts the MMM insensitive to the superconducting phases.

We have shown that MMM in the TJJ is created when an odd number of MZMs are brought into contact and coupled to each other.
Moreover, our analysis is based on an effective Hamiltonian, which focuses only on the couplings among the MZMs and is irrelevant to the details of the parent TS.
Thus, we expect that our strategy for creating a MMM can be extended to other systems,
such as higher-order TSs exhibiting high controllability for the number and location of MZMs~\cite{volovik_10,zhu_18,wang_18,ikegaya_21}.
Inspired by the various studies on the conventional MZMs, 
studying the Cooper pair splittings due to the nonlocal Andreev reflections~\cite{beenakker_08,law_13,law_14,akhmerov_18,flensberg_20,soori_20},
the nonlocal Josephson effects forming  Cooper quartets~\cite{melin_11,shtrikman_18,girit_19},
the quantum teleportation~\cite{fu_10,egger_11,egger_12},
and the full-counting statistics~\cite{schmidt_14,lutchyn_15,beenakker_15(2)} in the presence of MMMs would be intriguing topics for future works.
We hope that our work will stimulate further investigations into the systems hosting MMMs.

In summary, we demonstrate the appearance of the MMM whose wave function splits into the multiple parts localized at different edges of different TSs.
Thanks to the coexistence of the Majorana property and the nonlocal property, the MMM causes the drastic nonlocal resonant transport phenomena.
We expect that our proposal can be used to prove the realization of the fully functional topological TJJ, which forms an elemental qubit for fault-tolerant topological quantum computers.

\begin{acknowledgments}
S.I. is supported by the Grant-in-Aid for JSPS Fellows (JSPS KAKENHI Grant No. JP22KJ1507).
\end{acknowledgments}

\clearpage

\onecolumngrid
\begin{center}
  \textbf{\large Supplemental Material for ``Multi-locational Majorana Zero Modes''}\\ \vspace{0.3cm}
Yutaro Nagae$^{1}$, Andreas P. Schnyder$^{2}$, Yukio Tanaka$^{1}$, Yasuhiro Asano$^{3}$, and Satoshi Ikegaya$^{1,4}$\\ \vspace{0.1cm}
{\itshape $^{1}$Department of Applied Physics, Nagoya University, Nagoya 464-8603, Japan\\
$^{2}$Max-Planck-Institut f\"ur Festk\"orperforschung, Heisenbergstrasse 1, D-70569 Stuttgart, Germany\\
$^{3}$Department of Applied Physics, Hokkaido University, Sapporo 060-8628, Japan\\
$^{4}$Institute for Advanced Research, Nagoya University, Nagoya 464-8601, Japan}
\date{\today}
\end{center}

\section{Low-energy effective Hamiltonian}\label{sec:heff}

\begin{figure}[t]
\begin{center}
\includegraphics[width=0.75\textwidth]{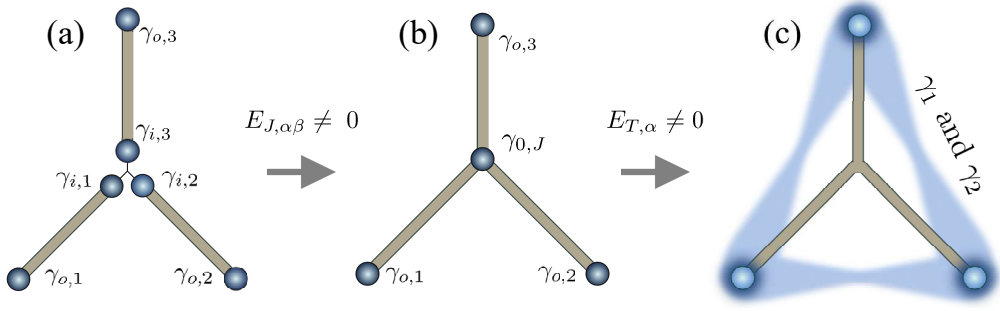}
\caption{Schematic image for the process creating Multi-locational Majorana zero modes.}
\label{fig:figure1sm}
\end{center}
\end{figure}

In this section, we derive the low-energy effective Hamiltonian for the three-terminal Josephson junctions consisting of the one-dimensional topological superconductors (TSs),
which is given by Eq.~(3) in the main text.
We start with the Hamiltonian in Eq.~(1) in the main text, which describes the couplings between Majorana zero modes (MZMs) in the three-terminal Josephson junction:
\begin{align}
\begin{split}
&H = \frac{1}{2} \boldsymbol{\gamma}^{\mathrm{T}} \mathcal{H} \boldsymbol{\gamma},\\
&\boldsymbol{\gamma}=[\gamma_{o,1},\gamma_{o,2},\gamma_{o,3},\gamma_{i,1},\gamma_{i,2},\gamma_{i,3}]^{\mathrm{T}},\quad
\mathcal{H}=\left( \begin{array}{cc} 0_{3\times3} & i \mathcal{H}_T \\ - i \mathcal{H}_T & \mathcal{H}_J \end{array} \right),\\
&\mathcal{H}_T = \mathrm{diag}[E_{T,1},E_{T,2},E_{T,3}],\quad
\mathcal{H}_{J} = \left( \begin{array}{ccc} 0 & i E_{J,12} & i E_{J,13} \\  -i E_{J,12} & 0 & i E_{J,23}\\ -i E_{J,13} & -i E_{J,23} & 0 \end{array} \right).
\label{eq:ham}
\end{split}
\end{align}
where $\gamma_{o(i), \alpha}=\gamma^{\dagger}_{o(i), \alpha}$ is an outer (inner) MZM of the $\alpha$-th TS.
In the absence of any couplings, as also illustrated in Fig.~\ref{fig:figure1sm}(a), one MZM appears at each end of the three TSs.
The coupling constant between the outer and inner MZMs in the $\alpha$-th TS is given by $E_{T,\alpha}$.
The coupling between the inner MZMs of the $\alpha$-th and $\beta$-th TSs at the Josephson junction is represented by $E_{J, \alpha\beta}=-E_{J,\beta\alpha}$.
The $n \times m$ null matrix is given by $0_{n\times m}$.
At first, we define a $3\times3$ unitary matrix that diagonalizes $\mathcal{H}_{J}$:
\begin{align}
\begin{split}
&\mathcal{U}_J^\dagger \mathcal{H}_J \mathcal{U}_J = \left(\begin{array}{cc}0 & 0_{1\times2} \\ 0_{2\times1} & \mathcal{E}_J\end{array}\right), \\
&\mathcal{E}_J=\left(\begin{array}{cc}E_J & 0 \\ 0 & -E_J\end{array} \right),\quad
E_J = \sqrt{E_{J,12}^2 + E_{J,23}^2 + E_{J,31}^2},
\end{split}
\end{align}
with
\begin{align}
\mathcal{U}_J = \left(\boldsymbol{V}_{0}, \boldsymbol{V}_{+}, \boldsymbol{V}_{-} \right),
\end{align}
where $\boldsymbol{V}_{0}$ and $\boldsymbol{V}_{\pm}$ are the three-element vectors satisfying
\begin{align}
\mathcal{H}_J \boldsymbol{V}_{0} = 0, \qquad
\mathcal{H}_J \boldsymbol{V}_{\pm} = \pm E_J\boldsymbol{V}_{\pm},
\end{align}
respectively.
Specifically, $\boldsymbol{V}_{0}$ is given by
\begin{align}
\boldsymbol{V}_{0} = \frac{1}{E_J}\left(\begin{array}{cl}E_{J,23} \\ -E_{J,13} \\ E_{J,12}\end{array} \right).
\end{align}
In addition, we define the field operators of
\begin{align}
\left(\begin{array}{cl} \gamma_{0,J}\\ \gamma_{+,J}\\ \gamma_{-,J}\end{array} \right) =
\mathcal{U}_J^{\dagger}\left(\begin{array}{cl} \gamma_{i,1} \\ \gamma_{i,2} \\ \gamma_{i,3}\end{array} \right)
\end{align}
where $\gamma_{0,J}$ and $\gamma_{\pm,J}$ corresponding to the field operators for the zero-energy eigenstate and the finite-energy eigenstates, respectively.
We note that $\gamma_{0,J}$ still satisfies the Majorana relation, $\gamma_{0,J} = \gamma_{0,J}^{\dagger}$.
By using the $6\times6$ unitary matrix of
\begin{align}
\mathcal{U} = \left(\begin{array}{cc}1_{3} & 0_{3\times3} \\ 0_{3\times3} & \mathcal{U}_J\end{array}\right),
\label{eq:unitary}
\end{align}
with $1_{n}$ being the $n \times n$ identity matrix, we rewrite the Hamiltonian as
\begin{align}
H = \frac{1}{2} \boldsymbol{\gamma}^{\mathrm{T}}\mathcal{U} \mathcal{U}^{\dagger} \mathcal{H} \mathcal{U} \mathcal{U}^{\dagger} \boldsymbol{\gamma}
= \frac{1}{2}\tilde{\boldsymbol{\gamma}}^{\dagger} \tilde{\mathcal{H}} \tilde{\boldsymbol{\gamma}},
\end{align}
with
\begin{align}
\tilde{\boldsymbol{\gamma}} =\mathcal{U}^{\dagger} \boldsymbol{\gamma} = [\gamma_{o,1},\gamma_{o,2},\gamma_{o,3},\gamma_{0,J},\gamma_{+,J},\gamma_{-,J}]^{T},
\end{align}
and
\begin{align}
\tilde{\mathcal{H}} = \mathcal{U}^{\dagger} \mathcal{H} \mathcal{U}
= \left(\begin{array}{cc} 0_{3\times3} & i\mathcal{H}_T \mathcal{U}_J \\ 
-i\mathcal{U}_J^\dagger\mathcal{H}_T & \mathcal{U}_J^\dagger\mathcal{H}_J \mathcal{U}_J \end{array}\right).
\end{align}
For later convenience, we rewrite the Hamiltonian as
\begin{align}
\begin{split}
&H = \frac{1}{2} \tilde{\boldsymbol{\gamma}}^{\dagger} 
\left( \begin{array}{cc} \mathcal{H}_0 & \mathcal{V} \\ \mathcal{V}^{\dagger} & \mathcal{E}_J \end{array} \right) \tilde{\boldsymbol{\gamma}},\\
&\mathcal{H}_0 = \left( \begin{array}{cccc} 0 & 0 & 0& iv_{1} \\ 0 & 0 & 0 & i v_{2} \\  0 & 0 & 0 & i v_{3}\\ -iv_{1} & -i v_{2} & -i v_{3} & 0 \end{array} \right),\quad
\mathcal{V}=\left(\begin{array}{cc} i v_{1,+} & iv_{1,-} \\ iv_{2,+} & iv_{2,-} \\ iv_{3,+} & iv_{3,-} \\ 0 & 0\end{array} \right),
\end{split}
\label{eq:ham2}
\end{align}
where
\begin{align}
\left(\begin{array}{cl} v_{1} \\ v_{2} \\ v_{3} \end{array} \right) = \mathcal{H}_T \boldsymbol{V}_{0} 
=  \frac{1}{E_J}\left(\begin{array}{cl}E_{T,1} E_{J,23} \\ -E_{T,2} E_{J,13} \\ E_{T,3} E_{J,12}\end{array} \right),
\end{align}
and
\begin{align}
\left(\begin{array}{cl} v_{1,\pm} \\ v_{2,\pm} \\ v_{3,\pm} \end{array} \right) = \mathcal{H}_T \boldsymbol{V}_{\pm}. 
\end{align}
The Hamiltonian in Eq.~(\ref{eq:ham2}) is equivalent to that in Eq.~(2) of the main text.
To construct the low-energy effective Hamiltonian, we consider the resolvent corresponding to $\tilde{\mathcal{H}}$~\cite{fu_18,timm_18}:
\begin{align}
\mathcal{G}(E) = (E-\tilde{\mathcal{H}})^{-1} =
\left( \begin{array}{cc} \mathcal{G}_{\mathrm{mzm}} & \mathcal{G}_{\mathcal{V}} \\ \mathcal{G}_{\mathcal{V}^{\dagger}} & \mathcal{G}_{J} \end{array} \right).
\end{align}
The block component related to the MZMs of
\begin{align}
\boldsymbol{\gamma}_{\mathrm{eff}} = [\gamma_{o,1},\gamma_{o,2},\gamma_{o,3},\gamma_{0,J}]^{T}
\end{align}
is obtained as
\begin{align}
\begin{split}
&\mathcal{G}_{\mathrm{mzm}}(E) = [E-\mathcal{H}_{\mathrm{mzm}}(E)]^{-1},\\
&\mathcal{H}_{\mathrm{mzm}}(E) = \mathcal{H}_0+\mathcal{V} (E-\mathcal{E}_J )^{-1}\mathcal{V}^{\dagger}.
\end{split}
\end{align}
When we focus on the low-energy excitation of $|E| \ll E_J$, we can make the replacement of
\begin{align}
\mathcal{V} (E-\mathcal{E}_J )^{-1}\mathcal{V}^{\dagger} \approx -\frac{1}{E_J} \mathcal{V} \left( \begin{array}{cc} 1&0\\0&-1 \end{array} \right) \mathcal{V}^{\dagger},
\end{align}
which yields the effective Hamiltonian describing the low-energy excitation of $|E| \ll E_J$~\cite{fu_18,timm_18}:
\begin{align}
\mathcal{H}_{\mathrm{eff}} = \mathcal{H}_0 + \delta \mathcal{H},\qquad
\delta \mathcal{H} = -\frac{1}{E_J} \mathcal{V} \left( \begin{array}{cc} 1&0\\0&-1 \end{array} \right) \mathcal{V}^{\dagger}.
\end{align}
The finite matrix elements in $\mathcal{H}_0$ are of the order of $E_T$, 
while finite matrix elements in $\delta \mathcal{H}$ are of the order of $E_T^{2}/E_J$,
where $E_T=\sqrt{E_{T,1}^2+E_{T,2}^2+E_{T,3}^2}$.
Therefore, in the limit of $E_T \ll E_J$, we can neglect the insignificant terms of $O(E^2_T/E_J)$ and obtain
\begin{align}
\mathcal{H}_{\mathrm{eff}} \approx \mathcal{H}_0.
\end{align}
Eventually, the effective Hamiltonian describing low-energy excitation of $|E|\ll E_J$ in the limit of $E_T \ll E_J$ is obtained by
\begin{align}
\begin{split}
&H_{\mathrm{eff}} = \frac{1}{2} \boldsymbol{\gamma}_{\mathrm{eff}}^{T}\mathcal{H}_{\mathrm{eff}}\boldsymbol{\gamma}_{\mathrm{eff}},\quad
\mathcal{H}_{\mathrm{eff}} = \left( \begin{array}{cccc} 0 & 0 & 0& iv_{1} \\ 0 & 0 & 0 & i v_{2} \\  0 & 0 & 0 & i v_{3}\\ -iv_{1} & -i v_{2} & -i v_{3} & 0 \end{array} \right),
\end{split}
\end{align}
which is equivalent to the Hamiltonian in Eq.~(3) of the main text.
As also illustrated in Fig.~\ref{fig:figure1sm}(b), we focus on the couplings between four MZMs: three outer MZMs and one MZM remained at the Josephson junction.

The effective Hamiltonian $H_{\mathrm{eff}}$ is diagonalized as,
\begin{align}
\mathcal{U}_{\mathrm{eff}}^{\dagger} \mathcal{H}_{\mathrm{eff}}\mathcal{U}_{\mathrm{eff}}=\mathcal{E}_{\mathrm{eff}}=\mathrm{diag}[0,0,E_v,-E_v],\qquad
E_v = \sqrt{v_1^2+v_2^2+v_3^2},
\end{align}
with
\begin{align}
\begin{split}
&\mathcal{U}_{\mathrm{eff}} = \left( \boldsymbol{u}_{0,1}, \boldsymbol{u}_{0,2}, \boldsymbol{u}_+, \boldsymbol{u}_- \right),\\
&\boldsymbol{u}_{0,1} = \left(\begin{array}{cl} \sin^2 {\varphi}(1-\cos \theta) + \cos \theta \\ -\sin \varphi \cos \varphi (1-\cos \theta) \\ - \cos \varphi \sin \theta \\0 \end{array} \right),\quad
\boldsymbol{u}_{0,2} = \left(\begin{array}{cl}-\sin \varphi \cos \varphi (1-\cos \theta) \\ \cos^2 {\varphi}(1-\cos \theta) + \cos \theta \\ - \sin \varphi \sin \theta \\0 \end{array} \right),\quad
\boldsymbol{u}_{\pm} = \frac{1}{\sqrt{2}} \left(\begin{array}{cl}\cos \varphi \sin \theta \\ \sin \varphi \sin \theta \\ \cos \theta \\ \pm i \end{array} \right),
\end{split}
\end{align}
where
\begin{align}
\left(\begin{array}{cl} v_{1} \\ v_{2} \\ v_{3} \end{array} \right)
=E_v \left(\begin{array}{cl} \cos \varphi \sin \theta \\ \sin \varphi \sin \theta \\ \cos \theta \end{array} \right).
\end{align}
As a result, we obtain the two zero-energy states:
\begin{align}
\gamma_n = \boldsymbol{u}_{0,n}^{\dagger}\boldsymbol{\gamma}_{\mathrm{eff}} = \sum_{\alpha=1\text{-}3} c_{\alpha}^{n} \gamma_{o,\alpha} \quad (n=1,2),
\end{align}
where we represent $\boldsymbol{u}_{0,n} = \left( c_1^n, c_2^n, c_3^n, 0 \right)^{\mathrm{T}}$.
Importantly, $\gamma_n$ is described by the superposition of the outer end MZMs (i.e., $\gamma_{o,\alpha}$).
Moreover, since $c_{\alpha}^{n}$ are the real number, $\gamma_n$ obeys $\gamma_{n}=\gamma_{n}^{\dagger}$.
Therefore, $\gamma_n$ describes the multi-locational Majorana zero modes (MMMs) as also illustrated in Fig.~\ref{fig:figure1sm}(c).
Strictly speaking, owing to the degeneracy at zero energy, there is ambiguity in the representation of the zero-energy eigenstates.
Nevertheless, the quantity of
\begin{align}
\begin{split}
&P_{\alpha} = \sum_{n=1,2}|c_{\alpha}^{n}|^2=\sum_{\beta\neq\alpha}\frac{v^2_{\beta}}{E_v^2},
\end{split}
\end{align}
which characterizes the amplitudes of the MZMs in the $\alpha$-th TS, is irrelevant to the representation of $c_{\alpha}^{n}$.

\section{Scattering coefficients}
In this section, we calculate the scattering coefficients in the presence of the MMM.
We start with the scattering matrix at the energy $E$, which is given in Eq.~(8) of the main text~\cite{beenakker_08,law_13,flensberg_20}:
\begin{align}
\begin{split}
&S(E)  = \left( \begin{array}{cc}s^{ee} & s^{eh} \\ s^{he} & s^{hh} \end{array} \right)
=1_6 - 2 \pi i W^\dagger \left[ E - \mathcal{H} + i \pi WW^\dagger \right]^{-1} W,\\
&W=\left( \begin{array}{cc} w & -w \\ 0_{3\times3} & 0_{3\times3} \end{array} \right), \quad
w=\left( \begin{array}{ccc} w_1 & 0 & 0 \\ 0 & w_2 & 0 \\ 0 & 0 & w_3 \end{array} \right), 
\end{split}
\end{align}
where the real number of $w_{\alpha}$ characterizes the coupling between the electron/hole in the $\alpha$-th lead and the MZM at the outer end of the $\alpha$-th TS.
By using the $6\times6$ unitary matrix in Eq.~(\ref{eq:unitary}), we can rewrite the scattering matrix as
\begin{align}
\begin{split}
S(E)  &= 1_6 - 2 \pi i W^\dagger \mathcal{U} \left[ E - \mathcal{U}^{\dagger}\mathcal{H}\mathcal{U} + i \pi \mathcal{U}^{\dagger}WW^\dagger\mathcal{U} \right]^{-1} \mathcal{U}^{\dagger}W\\
&= 1_6 - 2 \pi i W^\dagger \left[ E - \mathcal{\tilde{H}} + i \pi WW^\dagger \right]^{-1} W\\
&=1_6 - 2 \pi i W^\dagger \left( \begin{array}{cc} E - \mathcal{H}_{\mathrm{eff}} + i \pi \tilde{W}\tilde{W}^\dagger & -\mathcal{V}
\\ -\mathcal{V}^{\dagger} & E - \mathcal{E}_J \end{array} \right)^{-1}W\\
&=1_6 - 2 \pi i W^\dagger \left( \begin{array}{cc} \tilde{\mathcal{G}}_{\mathrm{mzm}} & \tilde{\mathcal{G}}_{\mathcal{V}} \\
\tilde{\mathcal{G}}_{\mathcal{V}^{\dagger}} & \tilde{\mathcal{G}}_{J} \end{array} \right)W\\
&=1_6 - 2 \pi i \tilde{W}^\dagger \tilde{\mathcal{G}}_{\mathrm{mzm}} \tilde{W},
\end{split}
\end{align}
where we use $\mathcal{U}^{\dagger}W=W$ and define,
\begin{align}
\begin{split}
&\left( \begin{array}{cc} \tilde{\mathcal{G}}_{\mathrm{mzm}} & \tilde{\mathcal{G}}_{\mathcal{V}} \\
\tilde{\mathcal{G}}_{\mathcal{V}^{\dagger}} & \tilde{\mathcal{G}}_{J} \end{array} \right)
=\left( \begin{array}{cc} E - \mathcal{H}_{\mathrm{eff}} + i \pi  \tilde{W}\tilde{W}^\dagger & -\mathcal{V} \\ -\mathcal{V}^{\dagger} & E - \mathcal{E}_J \end{array} \right)^{-1}, \\
&\tilde{W} =  \left( \tilde{w}, -\tilde{w} \right), \quad
\tilde{w}= \left( \begin{array}{cl} w \\ 0_{1\times3} \end{array} \right).
\end{split}
\end{align}
We specifically obtain
\begin{align}
\begin{split}
&\tilde{\mathcal{G}}_{\mathrm{mzm}}=\left[E - \mathcal{H}_{\mathrm{eff}} + i\pi  \tilde{W}\tilde{W}^\dagger - \mathcal{V} (E-\mathcal{E}_J )^{-1}\mathcal{V}^{\dagger}\right]^{-1}
=\tilde{\mathcal{G}}_{\mathrm{eff}} \left[1-\mathcal{V} (E-\mathcal{E}_J )^{-1}\mathcal{V}^{\dagger} \tilde{\mathcal{G}}_{\mathrm{eff}}\right]^{-1},\\
&\tilde{\mathcal{G}}_{\mathrm{eff}} = \left[E - \mathcal{H}_{\mathrm{eff}}  + i \pi \tilde{W}\tilde{W}^\dagger \right]^{-1}
= \left[E - \mathcal{H}_{\mathrm{eff}}  + i \tilde{\Gamma} \right]^{-1},\\
&\tilde{\Gamma}=\mathrm{diag}[\tilde{\Gamma}_1,\tilde{\Gamma}_2,\tilde{\Gamma}_3,0], \quad \tilde{\Gamma}_{\alpha} = 2\pi w_{\alpha}^2.
\end{split}
\end{align}
Similarly to the discussion in Sec.~\ref{sec:heff}, in the limit of $E, E_T, \tilde{\Gamma}_{\alpha} \ll E_J$, we can make the replacement of 
\begin{align}
\tilde{\mathcal{G}}_{\mathrm{mzm}} \approx \tilde{\mathcal{G}}_{\mathrm{eff}},
\end{align}
where the contributions from the finite-energy eigenstates at the Josephson junction, described by $\gamma_{\pm,J}$, are ignored.
Therefore, in the presence of the MMM, the low-energy scattering matrix is evaluated by
\begin{align}
S(E) =1_6 - 2 \pi i \tilde{W}^\dagger \tilde{\mathcal{G}}_{\mathrm{eff}} \tilde{W}
=1_6 - 2 \pi i \tilde{W}^\dagger \left[E - \mathcal{H}_{\mathrm{eff}}  + i \tilde{\Gamma} \right]^{-1} \tilde{W}.
\label{eq:smatfull}
\end{align}
From Eq.~(\ref{eq:smatfull}), we obtain the scattering coefficients at zero energy as
\begin{align}
\begin{split}
&s^{ee}_{\alpha\alpha} = \frac{v_{\alpha}^2 w_{\beta}^2 w_{\gamma}^2}{v_1^2w_2^2w_3^2+v_2^2w_3^2w_1^2+v_3^2w_1^2w_2^2},\quad
s^{he}_{\alpha\alpha} = -\frac{w_{\alpha}^2(v_{\beta}^2 w_{\gamma}^2+w_{\beta}^2 v_{\gamma}^2)}{v_1^2w_2^2w_3^2+v_2^2w_3^2w_1^2+v_3^2w_1^2w_2^2},\\
&s^{ee}_{\alpha\beta}=-s^{he}_{\alpha\beta}=\frac{v_{\alpha}v_{\beta}w_{\alpha}w_{\beta}w^{2}_{\gamma}}{v_1^2w_2^2w_3^2+v_2^2w_3^2w_1^2+v_3^2w_1^2w_2^2}, \quad
(\alpha\neq\beta,\;\beta\neq\gamma,\;\gamma\neq\alpha)
\end{split}
\end{align}
which is equivalent to Eq.~(10) in the main text.

\section{Numerical results on zero-frequency noise power}
\begin{figure}[h]
\begin{center}
\includegraphics[width=0.35\textwidth]{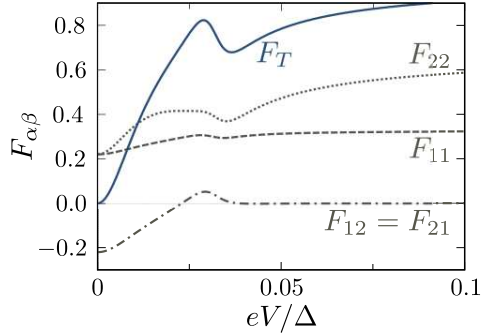}
\caption{Zero-frequency noise power normalized by the charge current as a function of the bias voltage.
The solid line represents $F_T=F_{11}+F_{22}+F_{12}+F_{21}$.}
\label{fig:figure2sm}
\end{center}
\end{figure}

In this section, we show the numerical results on the zero-frequency noise power calculated by using the tight-binding model,
where the scattering coefficients are computed using the lattice Green's function techniques~\cite{fisher_81,ando_91}.
We evaluate the zero-frequency noise power at zero temperature by~\cite{beenakker_94,datta_96}
\begin{align}
\begin{split}
&P_{\alpha\beta} = \frac{e^2}{h}\int^{eV}_{0} \mathcal{P}_{\alpha\beta}(E) dE,\\
&\mathcal{P}_{\alpha\beta}(E) = \delta_{\alpha\beta}\sum_{\eta=e,h}p^{\eta\eta}_{\alpha\alpha}
- \sum_{\eta,\zeta}\sigma_{\eta}\sigma_{\zeta}p^{\eta\zeta}_{\alpha\beta}p^{\zeta\eta}_{\beta\alpha},\quad
p^{\eta\zeta}_{\alpha}=\sum_{\gamma=1,2,3}s^{\eta e}_{\alpha \gamma} (s^{\zeta e}_{\beta\gamma})^{\ast},
\end{split}
\end{align}
where $\sigma_{\eta}=1$ ($-1$) for $\eta=e$ ($h$), which is equivalent to Eq.~(15) of the main text.
In Fig.~\ref{fig:figure2sm}, we show the zero-frequency noise power $P_{\alpha\beta}$ for $\alpha,\beta=1,2$ as a function of the bias voltage.
The parameters are chosen to be the same as those in Fig.~2(b) of the main text.
Specifically, we plot
\begin{align}
F_{\alpha\beta}(eV) = \frac{P_{\alpha\beta}(eV)}{I_1(eV)+I_2(eV)},
\end{align}
where $I_{\alpha}(eV)$ is the time-averaged current in the $\alpha$-th lead wire.
At the zero-bias voltage limit, we clearly find the relations of
\begin{align}
2|P_{12}|=P_{11}+P_{22}, \qquad
P_{11}+P_{22}+P_{12}+P_{21}=0,
\end{align}
which agree with Eq.~(17) and Eq.~(18) in the main text, respectively.
As a result, we numerically confirm the anomalous zero-frequency noise power due to the nonlocal resonant scatterings of the MMM.


\begin{thebibliography}{}
\bibitem{wilczek_09} F. Wilczek,
\href{https://www.nature.com/articles/nphys1380}{Nat. Phys. \textbf{5}, 614-618 (2009)}.
\bibitem{kane_10} M. Z. Hasan and C. L. Kane,
\href{https://journals.aps.org/rmp/abstract/10.1103/RevModPhys.82.3045}{Rev. Mod. Phys. \textbf{82}, 3045 (2010)}.
\bibitem{zhang_11}  X.-L. Qi  and  S.-C. Zhang,
\href{https://journals.aps.org/rmp/abstract/10.1103/RevModPhys.83.1057}{Rev. Mod. Phys. \textbf{83}, 1057 (2011)}.
\bibitem{schnyder_16} C.-K.Chiu, J. C. Y. Teo, A. P. Schnyder, and S. Ryu,
\href{https://journals.aps.org/rmp/abstract/10.1103/RevModPhys.88.035005}{Rev. Mod. Phys. \textbf{88}, 035005 (2016)}.
\bibitem{flensberg_12} M. Leijnse and K. Flensberg,
\href{https://iopscience.iop.org/article/10.1088/0268-1242/27/12/124003}{Semicond. Sci. Technol. \textbf{27}, 124003 (2012)}.
\bibitem{nagaosa_12} Y. Tanaka, M. Sato, and N. Nagaosa,
\href{https://journals.jps.jp/doi/abs/10.1143/JPSJ.81.011013}{J. Phys. Soc. Jpn. \textbf{81}, 011013 (2012)}.
\bibitem{beenakker_13} C. W. J. Beenakker,
\href{https://www.annualreviews.org/doi/abs/10.1146/annurev-conmatphys-030212-184337}{Annu. Rev. Condens. Matter Phys. \textbf{4}, 113-136 (2013)}.
\bibitem{sato_16} M. Sato and S. Fujimoto,
\href{https://journals.jps.jp/doi/10.7566/JPSJ.85.072001}{J. Phys. Soc. Jpn. \textbf{85}, 072001 (2016)}.
\bibitem{sato_17}  M. Sato and Y. Ando,
\href{https://iopscience.iop.org/article/10.1088/1361-6633/aa6ac7}{Rep. Prog. Phys. \textbf{80}, 076501 (2017)}.
\bibitem{kitaev_01} A. Y. Kitaev,
\href{https://iopscience.iop.org/article/10.1070/1063-7869/44/10S/S29}{Phys. Usp. \textbf{44}, 131 (2001)}.
\bibitem{volovik_97} G. E. Volovik,
\href{https://link.springer.com/article/10.1134/1.567563}{JETP Lett. \textbf{66}, 522 (1997)}.
\bibitem{green_00} N. Read and D. Green,
\href{https://journals.aps.org/prb/abstract/10.1103/PhysRevB.61.10267}{Phys. Rev. B \textbf{61}, 10267 (2000)}.
\bibitem{furusaki_01} A. Furusaki, M. Matsumoto, and M. Sigrist,
\href{https://journals.aps.org/prb/abstract/10.1103/PhysRevB.64.054514}{Phys .Rev. B \textbf{64}, 054514 (2001)}.
\bibitem{maiti_06} K. Sengupta, R. Roy, and M. Maiti,
\href{https://journals.aps.org/prb/abstract/10.1103/PhysRevB.74.094505}{Phys .Rev. B \textbf{74}, 094505 (2006)}.
\bibitem{schnyder_08} A. P. Schnyder, S. Ryu, A. Furusaki, and A. W. W. Ludwig,
\href{https://journals.aps.org/prb/abstract/10.1103/PhysRevB.78.195125}{Phys. Rev. B \textbf{78}, 195125 (2008)}.
\bibitem{zhang_09} X.-L. Qi, T. L. Hughes, S. Raghu, and S.-C. Zhang,
\href{https://journals.aps.org/prl/abstract/10.1103/PhysRevLett.102.187001}{Phys .Rev. Lett. \textbf{102}, 187001 (2009)}.
\bibitem{tanaka_09} Y. Tanaka, T. Yokoyama, A. V. Balatsky, and N. Nagaosa,
\href{https://journals.aps.org/prb/abstract/10.1103/PhysRevB.79.060505}{Phys .Rev. B \textbf{79}, 060505(R) (2009)}.
\bibitem{buchholtz_81} L. J. Buchholtz and G. Zwicknagl,
\href{https://journals.aps.org/prb/abstract/10.1103/PhysRevB.23.5788}{Phys. Rev. B \textbf{23}, 5788 (1981)}.
\bibitem{nagai_86} J. Hara and K. Nagai,
\href{https://academic.oup.com/ptp/article/76/6/1237/1891944}{Prog. Theor. Phys. \textbf{76}, 1237 (1986)}.
\bibitem{tanuma_01} Y. Tanuma, K. Kuroki, Y. Tanaka, and S. Kashiwaya,
\href{https://journals.aps.org/prb/abstract/10.1103/PhysRevB.64.214510}{Phys. Rev. B \textbf{64}, 214510 (2001)}.
\bibitem{sato_11} M. Sato, Y. Tanaka, K. Yada, and T. Yokoyama,
\href{https://journals.aps.org/prb/abstract/10.1103/PhysRevB.83.224511}{Phys. Rev. B \textbf{83}, 224511 (2011)}.
\bibitem{schnyder_11} A. P. Schnyder and S. Ryu,
\href{https://journals.aps.org/prb/abstract/10.1103/PhysRevB.84.060504}{Phys. Rev. B \textbf{84}, 060504(R) (2011)}.
\bibitem{balents_15} R.-J. Slager, L. Rademaker, J. Zaanen, and L. Balents,
\href{https://journals.aps.org/prb/abstract/10.1103/PhysRevB.92.085126}{Phys. Rev. B \textbf{92}, 085126 (2015)}.
\bibitem{hughes_17} W. A. Benalcazar, J. C. Y. Teo, and T. L. Hughes,
\href{https://journals.aps.org/prb/abstract/10.1103/PhysRevB.89.224503}{Phys. Rev. B \textbf{89}, 224503 (2014)}.
\bibitem{fang_17} Z. Song, Z. Fang, and C. Fang,
\href{https://journals.aps.org/prl/abstract/10.1103/PhysRevLett.119.246402}{Phys. Rev. Lett. \textbf{119}, 246402 (2017)}.
\bibitem{brouwer_17} J. Langbehn, Y. Peng, L. Trifunovic, F. vonOppen, and P. W. Brouwer,
\href{https://journals.aps.org/prl/abstract/10.1103/PhysRevLett.119.246401}{Phys. Rev. Lett. \textbf{119}, 246401 (2017)}.
\bibitem{khalaf_18} E. Khalaf, 
\href{https://journals.aps.org/prb/abstract/10.1103/PhysRevB.97.205136}{Phys. Rev. B \textbf{97}, 205136 (2018)}.
\bibitem{volovik_10} G. E. Volovik,
\href{https://link.springer.com/article/10.1134/S0021364010040090}{JETP Lett.\textbf{91}, 201-205 (2010)}.
\bibitem{zhu_18} X. Zhu,
\href{https://journals.aps.org/prb/abstract/10.1103/PhysRevB.97.205134}{Phys. Rev. B \textbf{97}, 205134 (2018)}.
\bibitem{wang_18} Z. Yan, F. Song, and Z. Wang,
\href{https://journals.aps.org/prl/abstract/10.1103/PhysRevLett.121.096803}{Phys. Rev. Lett. \textbf{121}, 096803 (2018)}.
\bibitem{ikegaya_21}S. Ikegaya, W. B. Rui, D. Manske, and A. P. Schnyder
\href{https://journals.aps.org/prresearch/abstract/10.1103/PhysRevResearch.3.023007}{Phys. Rev. Research \textbf{3} 023007 (2021)}.
\bibitem{thouless_82} D. J. Thouless, M. Kohmoto, M. P. Nightingale, and M. den Nijs
\href{https://journals.aps.org/prl/abstract/10.1103/PhysRevLett.49.405}{Phys. Rev. Lett. \textbf{49}, 405 (1982)}
\bibitem{kohmoto_85} M. Kohmoto,
\href{https://www.sciencedirect.com/science/article/pii/0003491685901484?via\%3Dihub}{ Ann. Phys. \textbf{160}, 343 (1985)}.
\bibitem{hatsugai_02} S. Ryu and Y. Hatsugai,
\href{https://journals.aps.org/prl/abstract/10.1103/PhysRevLett.89.077002}{Phys. Rev. Lett. \textbf{89}, 077002 (2002)}.
\bibitem{tanaka_95} Y. Tanaka and S. Kashiwaya,
\href{https://journals.aps.org/prl/abstract/10.1103/PhysRevLett.74.3451}{Phys. Rev. Lett. \textbf{74}, 3451 (1995)}.
\bibitem{sarma_01} K. Sengupta,  I. \v{Z}uti\'{c}, H.-J. Kwon, V. M. Yakovenko, and S. DasSarma,
\href{https://journals.aps.org/prb/abstract/10.1103/PhysRevB.63.144531}{Phys. Rev. B \textbf{63}, 144531 (2001)}.
\bibitem{tanaka_04} Y. Tanaka and S. Kashiwaya,
\href{https://journals.aps.org/prb/abstract/10.1103/PhysRevB.70.012507}{Phys. Rev. B \textbf{70}, 012507 (2004)}.
\bibitem{tanaka_05(1)} Y. Tanaka, S. Kashiwaya, and T. Yokoyama,
\href{https://journals.aps.org/prb/pdf/10.1103/PhysRevB.71.094513}{Phys. Rev. B \textbf{71}, 094513 (2005)}.
\bibitem{law_09} K. T. Law, P. A. Lee, and T. K. Ng,
\href{https://journals.aps.org/prl/abstract/10.1103/PhysRevLett.103.237001}{Phys. Rev. Lett. \textbf{103}, 237001 (2009)}.
\bibitem{asano_13} Y. Asano and Y. Tanaka,
\href{https://journals.aps.org/prb/abstract/10.1103/PhysRevB.87.104513}{Phys. Rev. B \textbf{87}, 104513 (2013)}. 
\bibitem{ikegaya_15} S. Ikegaya, Y. Asano, and Y. Tanaka,
\href{https://journals.aps.org/prb/abstract/10.1103/PhysRevB.91.174511}{Phys. Rev. B \textbf{91}, 174511 (2015)}.
\bibitem{ikegaya_16(1)} S. Ikegaya, S.-I. Suzuki, Y. Tanaka, and Y. Asano,
\href{https://journals.aps.org/prb/abstract/10.1103/PhysRevB.94.054512}{Phys. Rev. B \textbf{94}, 054512 (2016)}.
\bibitem{tanaka_97} Y. Tanaka and S. Kashiwaya,
\href{https://journals.aps.org/prb/abstract/10.1103/PhysRevB.56.892}{Phys. Rev. B \textbf{56}, 892 (1997)}.
\bibitem{kwon_04} H. -J. Kwon, K. Sengupta, and V. M. Yakovenko, 
\href{https://link.springer.com/article/10.1140/epjb/e2004-00066-4}{Eur. Phys. J. B \textbf{37}, 349--361(2004)}.
\bibitem{asano_06(1)} Y. Asano, Y. Tanaka, and S. Kashiwaya,
\href{https://journals.aps.org/prl/abstract/10.1103/PhysRevLett.96.097007}{Phys. Rev. Lett. \textbf{96}, 097007 (2006)}.
\bibitem{ikegaya_16(2)} S. Ikegaya and Y. Asano,
\href{https://iopscience.iop.org/article/10.1088/0953-8984/28/37/375702/meta}{J. Phys.: Condens. Matter \textbf{28}, 375702 (2016)}.
\bibitem{beenakker_08} J. Nilsson, A. R. Akhmerov, and C. W. J. Beenakker,
\href{https://journals.aps.org/prl/abstract/10.1103/PhysRevLett.101.120403}{Phys. Rev. Lett. \textbf{101}, 120403 (2008)}.
\bibitem{law_13}J. Liu, F.-C. Zhang, and K. T. Law,
\href{https://journals.aps.org/prb/abstract/10.1103/PhysRevB.88.064509}{Phys. Rev. B \textbf{88}, 064509 (2013)}.
\bibitem{law_14}J. J. He, J. Wu, T.-P. Choy, X.-J. Liu, Y. Tanaka, and K. T. Law,
\href{https://www.nature.com/articles/ncomms4232}{Nat. Commun. \textbf{5}, 3232 (2014)}.
\bibitem{akhmerov_18}T. \"O. Rosdahl, A. Vuik, M. Kjaergaard, and A. R. Akhmerov,
\href{https://journals.aps.org/prb/abstract/10.1103/PhysRevB.97.045421}{Phys. Rev. B \textbf{97} 045421 (2018)}.
\bibitem{flensberg_20}J. Danon, A. B. Hellenes, E. B. Hansen, L. Casparis, A. P. Higginbotham, and K. Flensberg,
\href{https://journals.aps.org/prl/abstract/10.1103/PhysRevLett.124.036801}{Phys. Rev. Lett. \textbf{124}, 036801 (2020)}.
\bibitem{soori_20}A. Soori,
\href{https://www.sciencedirect.com/science/article/abs/pii/S0038109820305597?via\%3Dihub}{Solid State Commun. \textbf{322}, 114055 (2020)}.
\bibitem{pachos_10} C. Benjamin and J. K. Pachos,
\href{https://journals.aps.org/prb/abstract/10.1103/PhysRevB.81.085101}{Phys. Rev. B \textbf{81}, 085101 (2010)}.
\bibitem{soori_19} A. Soori,
\href{https://iopscience.iop.org/article/10.1088/1361-648X/ab3f73}{J. Phys.: Condens. Matter \textbf{31}, 505301 (2019) }.
\bibitem{ardonne_17}C. Sp{\aa}nsl\"{a}tt and E. Ardonne,
\href{https://iopscience.iop.org/article/10.1088/1361-648X/aa585d/meta}{J. Phys.: Condens. Matter \textbf{29}, 105602 (2017)}.
\bibitem{sen_18}O. Deb, K. Sengupta, and D. Sen,
\href{https://journals.aps.org/prb/abstract/10.1103/PhysRevB.97.174518}{Phys. Rev. B \textbf{97}, 174518 (2018)}.
\bibitem{martin_19}T. Jonckheere, J. Rech, A. Zazunov, R. Egger, A. Levy Yeyati, and T. Martin,
\href{https://journals.aps.org/prl/abstract/10.1103/PhysRevLett.122.097003}{Phys. Rev. Lett. \textbf{122}, 097003 (2019)}.
\bibitem{houzet_21}J. S. Meyer and M. Houzet,
\href{https://journals.aps.org/prb/abstract/10.1103/PhysRevB.103.174504}{Phys. Rev. B 103, 174504 (2021)}.
\bibitem{asano_20}K. Sakurai, M. T. Mercaldo, S. Kobayashi, A. Yamakage, S. Ikegaya, T. Habe, P. Kotetes, M. Cuoco, and Y. Asano,
\href{https://journals.aps.org/prb/abstract/10.1103/PhysRevB.101.174506}{Phys. Rev. B \textbf{101}, 174506 (2020)}.
\bibitem{melin_11}A. Freyn, Benoit Dou\c{c}ot, D. Feinberg, and R. M\'elin,
\href{https://journals.aps.org/prl/abstract/10.1103/PhysRevLett.106.257005}{Phys. Rev. Lett. \textbf{106}, 257005 (2011)}.
\bibitem{shtrikman_18}Y. Cohen, Y. Ronen, J.-H. Kang, M. Heiblum, D. Feinberg, R. M\'elin, and H. Shtrikman,
\href{https://www.pnas.org/doi/full/10.1073/pnas.1800044115}{Proc. Natl. Acad. Sci. U.S.A. \textbf{115}, 6991-6994 (2018)}.
\bibitem{girit_19} J.-D. Pillet, V. Benzoni, J. Griesmar, J.-L. Smirr, and \c{C}. \"O. Girit,
\href{https://pubs.acs.org/doi/10.1021/acs.nanolett.9b02686}{Nano Lett. \textbf{19}, 7138-7143 (2019)}.
\bibitem{schmidt_14}L. Weithofer, P. Recher, and T. L. Schmidt,
\href{https://journals.aps.org/prb/abstract/10.1103/PhysRevB.90.205416}{Phys. Rev. B \textbf{90}, 205416 (2014)}.
\bibitem{lutchyn_15}D. E. Liu, A. Levchenko, and R. M. Lutchyn,
\href{https://journals.aps.org/prb/abstract/10.1103/PhysRevB.92.205422}{Phys. Rev. B \textbf{92}, 205422 (2015)}.
\bibitem{beenakker_15(2)}N. V. Gnezdilov, B. van Heck, M. Diez, J. A. Hutasoit, and C. W. J. Beenakker,
\href{https://journals.aps.org/prb/abstract/10.1103/PhysRevB.92.121406}{Phys. Rev. B \textbf{92}, 121406(R) (2015)}.
\bibitem{fu_10}L. Fu,
\href{https://journals.aps.org/prl/abstract/10.1103/PhysRevLett.104.056402}{Phys. Rev. Lett. \textbf{104}, 056402 (2010)}.
\bibitem{egger_11}A. Zazunov, A. L. Yeyati, and R. Egger,
\href{https://journals.aps.org/prb/abstract/10.1103/PhysRevB.84.165440}{Phys. Rev. B \textbf{84}, 165440 (2011)}.
\bibitem{egger_12}R. H\"utzen, A. Zazunov, B. Braunecker, A. L. Yeyati, and R. Egger,
\href{https://journals.aps.org/prl/abstract/10.1103/PhysRevLett.109.166403}{Phys. Rev. Lett. \textbf{109}, 166403 (2012)}.
\bibitem{ivanov_01} D. A. Ivanov,
\href{https://journals.aps.org/prl/abstract/10.1103/PhysRevLett.86.268}{Phys. Rev. Lett. \textbf{86}, 268 (2001)}.
\bibitem{kitaev_03} A. Y. Kitaev,
\href{https://www.sciencedirect.com/science/article/abs/pii/S0003491602000180?via\%3Dihub}{Ann. Phys. \textbf{303}, 2-30 (2003)}.
\bibitem{sarma_08} C. Nayak, S. H. Simon, A. Stern, M. Freedman, and S. DasSarma,
\href{https://journals.aps.org/rmp/abstract/10.1103/RevModPhys.80.1083}{Rev. Mod. Phys. \textbf{80}, 1083(2008)}.
\bibitem{alicea_16} D. Aasen, M. Hell, R. V. Mishmash, A. Higginbotham, J. Danon, M. Leijnse, T. S. Jespersen, J. A. Folk, C. M. Marcus, K. Flensberg, and J. Alicea,
\href{https://journals.aps.org/prx/abstract/10.1103/PhysRevX.6.031016}{Phys. Rev. X \textbf{6}, 031016 (2016)}.
\bibitem{fisher_11} J. Alicea, Y. Oreg, G. Refael, F. v. Oppen, and M. P. A. Fisher,
\href{https://www.nature.com/articles/nphys1915}{Nat. Phys. \textbf{7}, 412-417(2011)}.
\bibitem{tewari_11} J. D. Sau, D. J. Clarke, and S. Tewari,
\href{https://journals.aps.org/prb/abstract/10.1103/PhysRevB.84.094505}{Phys. Rev. B \textbf{84}, 094505 (2011)}.
\bibitem{sarma_15} S. DasSarma, M. Freedman, and C. Nayak,
\href{https://www.nature.com/articles/npjqi20151}{npj Quantum Information \textbf{1}, 15001(2015)}.
\bibitem{sen_16} O. Deb, M. Thakurathi, and D. Sen,
\href{https://link.springer.com/article/10.1140/epjb/e2015-60846-1}{Eur. Phys. J. B \textbf{89}, 19 (2016)}.
\bibitem{sakurai_20}K. Sakurai, M. T. Mercaldo, S. Kobayashi, A. Yamakage, S. Ikegaya, T. Habe, P. Kotetes, M. Cuoco, and Y. Asano,
\href{https://journals.aps.org/prb/abstract/10.1103/PhysRevB.101.174506}{Phys. Rev. B \textbf{101}, 174506 (2020)}.
\bibitem{dagotto_23}B. Pandey, N. Kaushal, G. Alvarez, and E. Dagotto,
\href{https://arxiv.org/abs/2306.04081}{arXiv:2306.04081}.
\bibitem{stanescu_12}S. DasSarma, J. D. Sau, and T. D. Stanescu,
\href{https://journals.aps.org/prb/abstract/10.1103/PhysRevB.86.220506}{Phys. Rev. B \textbf{86}, 220506(R) (2012)}.
\bibitem{sm} See Supplemental Material at XXX for the detailed derivation for Eq.~(\ref{eq:heff}) and Eq.~(\ref{eq:smat}).
We also show the numerical results on the current shot noise.
\bibitem{fu_18}J. W. F. Venderbos, L. Savary, J. Ruhman, P. A. Lee, and L. Fu,
\href{https://journals.aps.org/prx/abstract/10.1103/PhysRevX.8.011029}{Phys. Rev. X \textbf{8}, 011029 (2018)}.
\bibitem{timm_18}P. M. R. Brydon, D. F. Agterberg, Henri Menke, and C. Timm,
\href{https://journals.aps.org/prb/abstract/10.1103/PhysRevB.98.224509}{Phys. Rev. B \textbf{98}, 224509 (2018)}.
\bibitem{stern_21}K. Flensberg, F. v. Oppen, and A. Stern,
\href{https://www.nature.com/articles/s41578-021-00336-6}{Nat. Rev. Mater. \textbf{6}, 944 (2021)}.
\bibitem{yacoby_23}A. Yazdani, F. v. Oppen, B. I. Halperin, A. Yacoby,
\href{https://www.science.org/doi/10.1126/science.ade0850}{Science \textbf{380}, eade0850 (2023)}.
\bibitem{sarma_10} R. M. Lutchyn, J. D. Sau, and S. DasSarma,
\href{https://journals.aps.org/prl/abstract/10.1103/PhysRevLett.105.077001}{Phys. Rev. Lett. \textbf{105}, 077001 (2010)}.
\bibitem{oreg_10} Y. Oreg, G. Refael, and F. von Oppen,
\href{https://journals.aps.org/prl/abstract/10.1103/PhysRevLett.105.177002}{Phys. Rev. Lett. \textbf{105}, 177002 (2010)}.
\bibitem{kouwenhoven_12} V. Mourik, K. Zuo, S. M. Frolov, S. R. Plissard, E. P. A. M. Bakkers, and L. P. Kouwenhoven,
\href{https://science.sciencemag.org/content/336/6084/1003}{Science \textbf{336}, 1003-1007 (2012)}.
\bibitem{deng_12} M. T. Deng, C. L. Yu, G. Y. Huang, M. Larsson, P. Caroff, and H. Q. Xu,
\href{https://pubs.acs.org/doi/10.1021/nl303758w}{Nano Lett. \textbf{12}, 6414-6419 (2012)}.
\bibitem{beenakker_11} T.-P. Choy, J. M. Edge, A. R. Akhmerov, and C. W. J. Beenakker,
\href{https://journals.aps.org/prb/abstract/10.1103/PhysRevB.84.195442}{Phys. Rev. B \textbf{84}, 195442 (2011)}.
\bibitem{yazdani_13} S. Nadj-Perge, I. K. Drozdov, B. A. Bernevig, and A. Yazdani,
\href{https://journals.aps.org/prb/abstract/10.1103/PhysRevB.84.195442}{Phys. Rev. B \textbf{88}, 020407(R) (2013)}.
\bibitem{yazdani_14} S. Nadj-Perge, I. K. Drozdov, J. Li, H. Chen, S. Jeon, J. Seo, A. H. MacDonald, B. A. Bernevig, and A. Yazdani,
\href{https://science.sciencemag.org/content/346/6209/602}{Science \textbf{346}, 602-607 (2014)}.
\bibitem{yazdani_17} B. E. Feldman, M. T. Randeria, J. Li, S. Jeon, Y. Xie, Z. Wang, I. K. Drozdov,  B. A. Bernevig, and A. Yazdani,
\href{https://www.nature.com/articles/nphys3947}{Nat. Phys. \textbf{13}, 286-291 (2017)}.
\bibitem{flensberg_17} M. Hell, M. Leijnse, and K. Flensberg,
\href{https://journals.aps.org/prl/abstract/10.1103/PhysRevLett.118.107701}{Phys. Rev. Lett. \textbf{118}, 107701 (2017)}.
\bibitem{halperin_17} F. Pientka, A. Keselman, E. Berg, A. Yacoby, A. Stern, and B. I. Halperin,
\href{https://journals.aps.org/prx/abstract/10.1103/PhysRevX.7.021032}{Phys. Rev. X \textbf{7}, 021032 (2017)}.
\bibitem{haim_19} A. Haim and A. Stern,
\href{https://journals.aps.org/prl/abstract/10.1103/PhysRevLett.122.126801}{Phys. Rev. Lett. \textbf{122}, 126801 (2019)}.
\bibitem{setiawan_19} F. Setiawan, A. Stern, and E. Berg,
\href{https://journals.aps.org/prb/abstract/10.1103/PhysRevB.99.220506}{Phys. Rev. B \textbf{99}, 220506(R) (2019)}.
\bibitem{nichele_19} A. Fornieri, A. M. Whiticar, F. Setiawan, E. Portol\'{e}s, A. C. C. Drachmann, A. Keselman, S. Gronin, C. Thomas, T. Wang, R. Kallaher,
G. C. Gardner, E. Berg, M. J. Manfra, A. Stern, C. M. Marcus, and F. Nichele,
\href{https://www.nature.com/articles/s41586-019-1068-8}{Nature \textbf{569}, 89-92 (2019)}.
\bibitem{yacoby_19} H. Ren, F. Pientka, S. Hart, A. T. Pierce, M. Kosowsky, L. Lunczer, R. Schlereth, B. Scharf, E. M. Hankiewicz, L. W. Molenkamp, B. I. Halperin, and A. Yacoby,
\href{https://www.nature.com/articles/s41586-019-1148-9}{Nature \textbf{569}, 93-98 (2019)}.
\bibitem{shabani_20} T. Zhou, M. C. Dartiailh, W. Mayer, J. E. Han, A. Matos-Abiague, J. Shabani, and I. \v{Z}uti\'{c},
\href{https://journals.aps.org/prl/abstract/10.1103/PhysRevLett.124.137001}{Phys. Rev. Lett. \textbf{124}, 137001 (2020)}.
\bibitem{shabani_21} M. C. Dartiailh, W. Mayer, J. Yuan, K. S. Wickramasinghe, A. Matos-Abiague, I. \v{Z}uti\'{c}, and J. Shabani,
\href{https://journals.aps.org/prl/abstract/10.1103/PhysRevLett.126.036802}{Phys. Rev. Lett. \textbf{126}, 036802 (2021)}.
\bibitem{ikegaya_22} D. Oshima, S. Ikegaya, A. P. Schnyder, and Y. Tanaka,
\href{https://journals.aps.org/prresearch/abstract/10.1103/PhysRevResearch.4.L022051}{Phys. Rev. Research \textbf{4}, L022051 (2022)}.
\bibitem{thouless_74} D. Thouless,
\href{https://www.sciencedirect.com/science/article/pii/0370157374900295?via\%3Dihub}{Phys. Rep. \textbf{13}, 93 (1974)}.
\bibitem{fisher_81} P. A. Lee and D. S. Fisher,
\href{https://journals.aps.org/prl/abstract/10.1103/PhysRevLett.47.882}{Phys. Rev. Lett. \textbf{47}, 882 (1981)}.
\bibitem{ando_91} T. Ando,
\href{https://journals.aps.org/prb/abstract/10.1103/PhysRevB.44.8017}{Phys. Rev. B \textbf{44}, 8017 (1991)}.
\bibitem{klapwijk_82} G. E. Blonder, M. Tinkham, and T. M. Klapwijk,
\href{https://journals.aps.org/prb/abstract/10.1103/PhysRevB.25.4515}{Phys .Rev. B \textbf{25}, 4515 (1982).}
\bibitem{deutscher_00} G. Deutscher and D. Feinberg,
\href{https://aip.scitation.org/doi/abs/10.1063/1.125796}{Appl. Phys. Lett. \textbf{76}, 487-489 (2000).}
\bibitem{beenakker_94}M. J. M. de Jong and C. W. J. Beenakker,
\href{https://journals.aps.org/prb/abstract/10.1103/PhysRevB.49.16070}{Phys. Rev. B \textbf{49}, 16070 (1994)}.
\bibitem{datta_96}M. P. Anantram and S. Datta,
\href{https://journals.aps.org/prb/abstract/10.1103/PhysRevB.53.16390}{Phys. Rev. B \textbf{53}, 16390 (1996)}.
\end{thebibliography}
\end{document}